\documentclass[conference]{IEEEtran}
\IEEEoverridecommandlockouts
\usepackage{cite}
\usepackage{amsmath,amssymb,amsfonts}
\usepackage{algorithmic}
\usepackage{graphicx}
\usepackage{textcomp}
\usepackage{xcolor}
\def\BibTeX{{\rm B\kern-.05em{\sc i\kern-.025em b}\kern-.08em
    T\kern-.1667em\lower.7ex\hbox{E}\kern-.125emX}}
\begin{document}

\bstctlcite{IEEEexample:BSTcontrol}

\title{Contactless seismocardiography via Gunnar-Farneback optical flow \\
\thanks{*This work was supported by the National Science Foundation under Grant No. 2340020 and a SMART Business Act Grant through Mississippi Institutions of Higher Learning (Grant No. 2024-04).}
}

\author{\IEEEauthorblockN{Mohammad Muntasir Rahman}
		\IEEEauthorblockA{\textit{Dept of Ag. \& Biological Engineering} \\
			\textit{Mississippi State University}\\
			Mississippi State, MS 39762, USA \\
			mmr510@msstate.edu}
		\and
		\IEEEauthorblockN{Amirtahà Taebi}\thanks{Corresponding author: ataebi@abe.msstate.edu}
		\IEEEauthorblockA{\textit{Dept of Ag. \& Biological Engineering} \\
			\textit{Mississippi State University}\\
			Mississippi State, MS 39762, USA \\
			ataebi@abe.msstate.edu}		
}
	
\maketitle

\begin{abstract}
Seismocardiography (SCG) has gained significant attention due to its potential applications in monitoring cardiac health and diagnosing cardiovascular conditions. Conventional SCG methods rely on accelerometers attached to the chest, which can be uncomfortable or inconvenient. In recent years, researchers have explored non-contact methods to capture SCG signals, and one promising approach involves analyzing video recordings of the chest. In this study, we investigate a vision-based method based on the Gunnar-Farneback optical flow to extract SCG signals from the chest skin movements recorded by a smartphone camera. We compared the SCG signals extracted from the chest videos of four healthy subjects with those obtained from accelerometers and our previous method based on sticker tracking. Our results demonstrated that the vision-based SCG signals extracted by the proposed method closely resembled those from accelerometers and stickers, although these signals were captured from slightly different locations. The mean squared error between the vision-based SCG signals and accelerometer-based signals was found to be within a reasonable range, especially between signals on head-to-foot direction (0.2$<$MSE$<$1.5). Additionally, heart rates derived from the vision-based SCG exhibited good agreement with the gold-standard ECG measurements, with a mean difference of 0.8 bpm. These results indicate the potential of this non-invasive method in health monitoring and diagnostics.

\end{abstract}


\section{Introduction}
Cardiovascular diseases have been a significant global health concern, imposing substantial burdens on economies worldwide \cite{tsao2022heart}. Monitoring cardiac health early in its progression is crucial for effective prevention and management. Seismocardiogram (SCG) is a valuable physiological signal that provides insights into cardiac function and health \cite{cook2022body}. Unlike traditional electrocardiography (ECG), which measures electrical activity, the SCG focuses on mechanical vibrations generated by the heart’s movements \cite{taebi2019recent}. These vibrations propagate through the chest wall and can be detected using accelerometers or other contact-based sensors \cite{taebi2019recent,rahman2023reconstruction}. However, sensor attachment can be uncomfortable or inconvenient for patients. Alternatively, technologies such as infrared sensors \cite{boccanfuso2012collecting}, radars \cite{nosrati2018high}, and WiFi devices \cite{gu2019wifi} are used to capture SCG signals without physical contact. Moreover, advancements in computer vision and motion tracking techniques have opened up new possibilities for non-contact SCG signal acquisition \cite{rahman2023non,que2022contactless}. In our prior work \cite{rahman2023non}, we introduced a vision-based approach for extracting right-to-left and head-to-foot SCG signals from chest videos by tracking a patterned sticker attached to the chest using the Lucas-Kanade method. Specifically, we utilized texture-patterned stickers, since they create a high-contrast artificial region allowing the algorithm to reliably identify and track them \cite{wang2015digital}. In this proof-of-concept study, we explore the feasibility of capturing chest vibrations and SCG signals from video recordings of the chest surface without using any attached patterned stickers. For this task, we define a small region of skin on the chest surface and employ the Farneback optical flow algorithm \cite{farneback2003two} to analyze that region for motion tracking in each consecutive frame. By tracking pixel displacements over time, we can infer the subtle vibrations associated with cardiac wall motion. This approach eliminates the need for attaching patterned stickers to the chest or placing sensors on it, making it more comfortable for patients and enabling continuous monitoring in real-world scenarios. Our findings contribute to the growing field of contactless health monitoring, emphasizing the feasibility of video-based SCG analysis for early detection of cardiovascular abnormalities.

\section{Materials and Methods}

\subsection{Study Population and Data Acquisition}
The study was approved by Mississippi State University's Institutional Review Board. Data was collected from four healthy human subjects (age: 26.3 $\pm$ 6.7 years; BMI: 25.8 $\pm$ 6.3 kg/m$^2$) after obtaining informed consent. 

\begin{figure}
	\centering
	\includegraphics[width=.8\columnwidth]{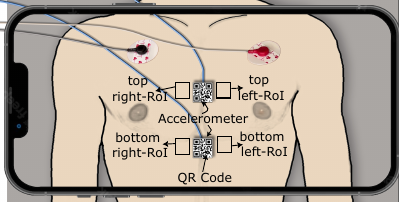}
	\caption{Data acquisition and sensor placement setup.}
	\label{fig:data_acquision}
\end{figure}

During data collection, subjects rested supine on a bed without additional body movements. To validate the vision-based SCG signals, we attached two triaxial accelerometers to the sternum, one next to the fourth costal notch and the other on the xiphoid process. These locations were referred to as the “top” and “bottom,” respectively. Simultaneously, single-lead ECG data were acquired using the same data acquisition system. Additionally, to compare the SCG signals captured from skin movements with those from a high-contrast and trackable sticker, we attached QR codes on top of each accelerometer as shown in Fig. \ref{fig:data_acquision}. An iPhone 13 Pro (Apple Inc., Cupertino, CA) was used to record chest motion videos at 60 fps with a resolution of $3840 \times 2160$ pixels. To keep the phone stationary, we used a phone holder with the back camera facing the subject’s chest. In addition, a Bluetooth remote control was employed to start and stop recording. To synchronize the accelerometer, ECG, and video data, we utilized a microphone connected to both the data acquisition system and the smartphone. The microphone was tapped at the beginning and end of each recording, and these timestamps were used to synchronize the video, accelerometer SCG, and ECG data during two 15-second breath-hold maneuvers: one at the end of inhalation (BHEI) and the other at the end of exhalation (BHEE).

\subsection{Vision-based SCG Signal Extraction}
After recording the video, the objective is to extract SCG signals from specific regions of interest (RoIs) on the skin. These RoIs were selected on the first frame of the video, corresponding to the onset of the first tap sound in the audio signal. For both top and bottom locations, two RoIs on the skin were selected, one to the left of the sensor and the other to the right, referred to as “left-RoI” and “right-RoI” (Fig. \ref{fig:data_acquision}). Also, the stickers' RoIs were defined based on the QR codes affixed to each accelerometer. After selecting the RoIs, their motion across subsequent frames of the video was tracked using the Gunnar-Farneback optical flow algorithm \cite{farneback2003two} which is a two-frame dense motion estimation technique, aiming at computing the motion of pixels between consecutive frames in a video sequence. Unlike sparse optical flow methods that track specific feature points, Farneback’s approach considers all points in the image. It leverages polynomial expansion to approximate the neighborhood of each image pixel. The algorithm constructs a pyramid of images, with each level having a lower resolution than the previous one. This pyramid helps handle motions of varying scales. At each level, the algorithm performs polynomial expansion to estimate the intensity changes between corresponding pixels in two consecutive frames. By minimizing the sum of squared differences between predicted and actual intensities, it iteratively searches for the best displacement at each pixel level. The final result is a high-resolution optical flow map consisting of a displacement vector ($dx/dt$, $dy/dt$) for each pixel.

In each video frame, displacement pairs ($dx, dy$) are obtained for all pixels inside the RoI in the right-to-left ($x$) and head-to-foot ($y$) directions. To compute the overall displacement for each RoI, we calculated the median displacement of all pixels within the RoI: $D=\text{median}\{d_1, d_2, ..., d_n\}$, where $\{d_1, d_2, ..., d_n\}$ represent the displacement values extracted from all the pixels within the RoI. After calculating the overall displacement pairs ($Dx,Dy$) for each RoI across the entire video, the second derivative of the displacement was determined to obtain the acceleration signal in the $x$ and $y$ directions that represent the vision-based SCG signal in right-to-left and head-to-foot directions.

\begin{figure*}[ht]
    \centering
    \includegraphics[width=\linewidth]{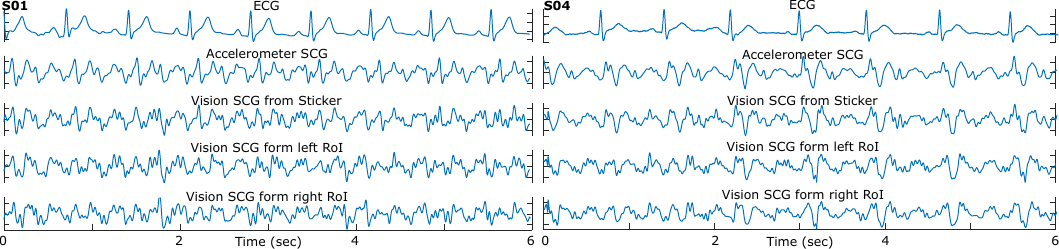}
    \caption{Comparison of the vision-based signals and the gold-standard accelerometer output. The signals are shown for subjects 1 and 4 (S01 and S04) from the bottom location recorded during breath-hold at the end of exhalation. Only the head-to-foot component of the SCG signals are shown.}
    \label{fig:scg_signals}
\end{figure*}

\subsection{Signal Denoising and Synchronization}
The SCG signals were detrended by removing the best straight-fit line from the data. Then, the accelerometer and vision-based SCG signals were filtered using a 4th-order Butterworth bandpass filter with cutoff frequencies of 1 and 30 Hz. The lower cutoff frequency was used to remove the respiratory noise. The higher cutoff frequency was selected as 30 Hz because the SCG estimations from the video could capture vibrations up to half of the camera acquisition speed, i.e., 60 fps. We then resampled the vision-based SCG signals using linear interpolation to 5000 Hz to match the sampling frequency of the accelerometer and ECG signals.

The vision-based and accelerometer SCG signals were recorded by two distinct systems. Although we synchronized the signals using audio taps at the beginning and end of the recordings, a slight lag was still present between the accelerometer SCG and the vision-based SCG captured from the stickers attached to the corresponding accelerometer. Additionally, during the resampling step, we applied an FIR antialiasing lowpass filter, which may introduce some delay. To remove this lag, we calculated the time differences between the sticker’s SCG and the corresponding accelerometer signal via cross-correlation, and removed them to sync the signals.

\begin{figure*}[ht!]
	\centering
    \includegraphics[width=1\linewidth]{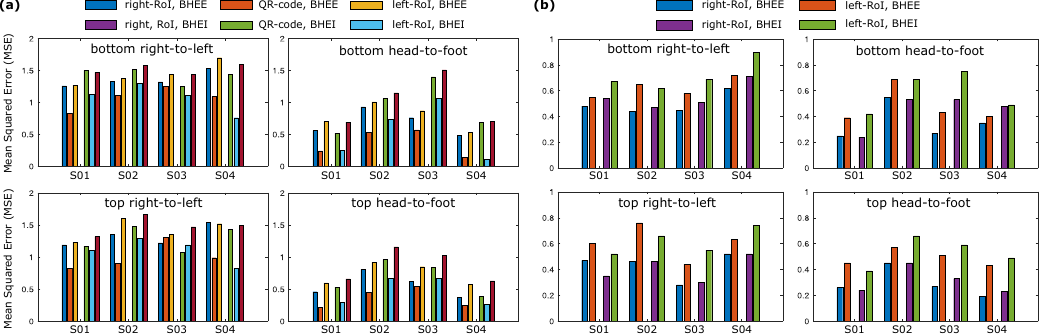}
	\caption{(a) MSE between vision-based and accelerometer signals, (b) MSE between vision-based signals extracted from the QR code sticker and the RoIs.}
	\label{fig:mse_error}
\end{figure*}

\begin{figure}[t]
	\centering
	\includegraphics[width=\columnwidth]{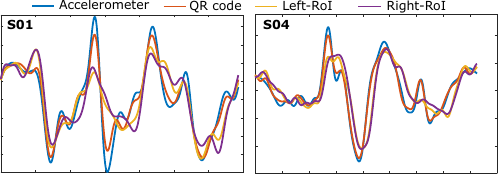}
	\caption{Signal segment derived from the SCG signal obtained at the bottom location, recorded at the end of exhalation in the head-to-foot direction.}
	\label{fig:scg_segment}
\end{figure}

\subsection{Heart Rate Estimation from Chest Videos}
We estimated heart rates (HRs) from the head-to-foot vision-based SCG signals. For this purpose, we further denoised the signals using some of the methods outlined in \cite{liu2024camera}. Specifically, we applied a Savitzky-Golay smoothing filter with a length of 60 ms to achieve signal smoothness. We then used a 4th-order Butterworth bandpass filter with cutoff frequencies of 1 Hz and 15 Hz. Then, we decomposed the SCG signals using the variational mode decomposition (VMD). The last two modal components with the lowest frequencies were then used to estimate HR. 

For this purpose, a narrower bandpass filter (0.75-1.5 Hz) was first applied to further isolate the heartbeats. Next, the peaks of this filtered signal were identified using a minimum peak distance of half the sampling frequency and a minimum peak prominence of 10\% of the filtered signal's standard deviation. The time interval between consecutive peaks was used to calculate the instantaneous HR. K-means clustering with two clusters was then employed to identify and remove outliers in the instantaneous HR data. The larger cluster's average was assumed to provide a first estimation of the HR, and any values deviating by more than 20\% from it were excluded from the instantaneous HR. Finally, the remaining HR values were averaged to obtain the mean HR.

\begin{figure}[t]
	\centering
	\includegraphics[width=\columnwidth]{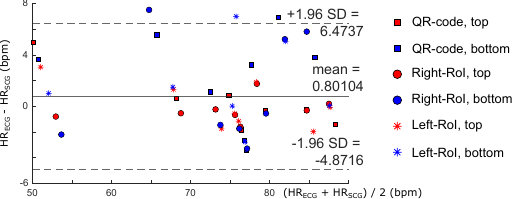}
	\caption{Agreement between heart rates (HRs) extracted from ECG and vision-based SCG.}
	\label{fig:hr_bland_altman}
\end{figure}

\section{Results and Discussion}
\subsection{Vision-based SCG}
Fig. \ref{fig:scg_signals} shows samples of the SCG signals in the head-to-foot direction along with the corresponding ECG signal for subjects 1 and 4 (S01 and S04). The three vision-based signals shown include the SCG signal extracted from the sticker on top of the accelerometer and the SCG signals extracted from the chest skin movements within the right- and left-RoIs. Previous studies have extensively documented intra-subject variability in SCG signals \cite{inan2014ballistocardiography, taebi2019recent}. Therefore, this might be the source of some differences in the signals obtained from these locations and the gold-standard accelerometer SCG.

\subsection{Signal Comparison}
Although the right- and left-RoIs are slightly away from the accelerometer, we compared the signals obtained from these locations with the corresponding accelerometer and sticker-based SCG. We visually observed a strong similarity between the accelerometer signal and the vision-based signal extracted from the sticker at each sensor location. Remarkably, despite the location differences, our visual inspection revealed that the SCG signals extracted from the right- and left-RoIs closely resembled the accelerometer signal. Furthermore, we quantified the signal similarity by calculating the mean squared error (MSE) between the vision-based SCG signals and the corresponding accelerometer signal in the right-to-left and head-to-foot directions (Fig. \ref{fig:mse_error}.a). The MSE between the sticker-based SCG and accelerometer signal (depicted in the middle of each block in the chart) was relatively lower, as these signals were captured from the same chest location despite recording by two different systems. However, when comparing the right- and left-RoI SCG signals with the accelerometer output, we observed a slightly higher MSE. This discrepancy can be attributed to the vision-based SCG being captured from locations slightly different than the sensor positions. Additionally, the SCG signals in the head-to-foot direction exhibited higher similarity than those in the right-to-left direction, resulting in a larger MSE for the latter. Furthermore, the MSE between the right- and left-RoI SCG signals and the sticker-based SCG signal is shown in Fig. \ref{fig:mse_error}.b. These results suggest that the right-RoI SCG signal consistently had a higher similarity with the sticker-based signal than the left-RoI signal. This observation may be attributed to the well-known variation in SCG signals across the chest, where locations closer to the cardiac structure experience more pronounced effects from cardiac vibrations. These findings suggest the feasibility of our proposed method in extracting SCG signals from the video recordings of the chest skin. However, we noticed that the quality of these signals is affected by the presence of natural chest landmarks. For example, the presence of chest hair can enhance the performance of the proposed method.

\subsection{SCG Variability}
To investigate SCG variability across measurement locations, we segmented the SCG signals into cardiac cycles using the ECG R peaks detected by the Pan-Tompkins algorithm. For each subject, the average heart cycle duration, $t_c$, was calculated based on the ECG RR intervals and was used to define the size of the SCG segments. Each segment was set to start from $t_c/4$ before the corresponding R peak to ensure consistent segmentation of the SCG signal across all subjects. Then we calculated the ensemble average of all segments for every SCG signal. Figure \ref{fig:scg_segment} illustrates the ensemble average of the head-to-foot SCG signal extracted from the bottom location at BHEE for S01 and S04. When comparing the accelerometer signal with the sticker (QR code) signal, a higher degree of similarity is observed. Although the signals were recorded using different systems, they were captured from the same location, which justifies their higher similarity. However, the skin-based SCG signal, captured from slightly different locations (left-RoI and right-RoI), exhibits deviations compared to the accelerometer and sticker-based signals, most likely due to different signal acquisition locations.

\subsection{Heart Rate Agreement Analysis}
We determined HRs using data from the left-RoI, right-RoI, and stickers at both top and bottom locations and compared them with the ECG-based HR. The Bland-Altman plot shows a good agreement between the HRs estimated from vision-based SCG and the gold-standard ECG (Fig. \ref{fig:hr_bland_altman}). The mean difference (bias) was 0.80 bpm, with lower and upper limits of agreement ranging from -4.87 to 6.47 bpm.

\section{Conclusion}
In this study, we investigated the feasibility of extracting SCG signals from the video recordings of the chest skin using Gunnar-Farneback optical flow algorithm. This vision-based method presents a contactless alternative to traditional SCG techniques that rely on accelerometers attached to the chest. Our findings indicate that SCG signals extracted from the chest videos closely resemble those obtained from accelerometers, and important cardiac parameters such as HR can be accurately estimated from these vision-based SCG signals. However, further research is necessary to enhance the robustness of this method on a larger and more diverse population.

\bibliographystyle{IEEEtran}	
\bibliography{references}	

\end{document}